\documentstyle[aps,pre]{revtex}
\begin{document}
\draft
\title{Tracer dispersion in two-dimensional rough fractures}
\author{German Drazer and Joel Koplik}
\address{Benjamin Levich Institute and Department of Physics,
City College of the City University of New York, New York, NY 10031}
\date{\today}
\maketitle
\begin{abstract}
Tracer diffusion and hydrodynamic dispersion in two-dimensional
fractures with self-affine roughness is studied by
analytic and numerical methods. 
Numerical simulations were performed via the lattice-Boltzmann
approach, using a new boundary condition for tracer particles 
that improves the accuracy of the method. 
The reduction in the diffusive transport, due to 
the fractal geometry of the fracture surfaces, 
is analyzed for different fracture apertures.
In the limit of small aperture fluctuations we
derive the correction to the diffusive coefficient 
in terms of the tortuosity, which accounts for 
the irregular geometry of the fractures.
Dispersion is studied when the two fracture surfaces are 
simple displaced normally to the mean fracture plane, and when
there is a lateral shift as well. Numerical results are analyzed
using the $\Lambda$-parameter, related to convective transport within
the fracture, and simple arguments based on
lubrication approximation. At very low P\'eclet number, in the case where 
fracture surfaces are laterally shifted, we show 
using several different methods that convective
transport reduces dispersion. 
\end{abstract}
\section{Introduction}
\label{introduction}

Tracer spreading in flows between parallel walls has received
considerable attention since the celebrated work
of Taylor\cite{taylor}. This problem is of fundamental importance
in a variety of fields, and in particular transport processes 
through artificial or natural porous media. In general, the 
dispersion of a tracer carried along by a flowing fluid in a 
medium of disordered structure, such as hydrocarbon or water
reservoirs, involves a combination of convection-dispersion through
the microscopic pore space of the rock itself and through macroscopic
channels such as fractures. In the simplest case of
Poiseuille flow in a Hele-Shaw cell, which in several 
applications is used to model fractures
\cite{adler99,bear93,sahimi95,nas}, 
the vanishing velocity near the solid gives rise to
a large dispersion, quadratic in the P\'eclet number
\cite{taylor,aris59}. However, one frequently encounters systems in
which the channel aperture varies along the flow direction, 
or in which the channel walls have some rugosity. This is the case
for fractured rocks, in which fractured boundaries
can be usually described as correlated,
self-affine fractals\cite{bouchaud97}. The roughness
exponent, is usually found to be close to $0.8$,
 insensitive to the material and fracturing process
\cite{bouchaud90,plouraboue96,boffa}. 
Most of the studies dealing with varying
channel aperture model the fluctuations
as slowly varying periodic functions \cite{sobey85}.
Only a few works deal with more realistic models
of fractures, such as random rugosities perpendicular to the flow
\cite{koplik93} or self-affine fractures
\cite{gutfraind95,plouraboue98}. 

In the present paper we present numerical simulations of
tracer diffusion and dispersion in self-affine 
fractures. The simulations are two-dimensional,
(two-dimensional fracture surfaces),  
and the fluid flow and tracer spreading are computed using 
the lattice-Boltzmann (LB) method. 
We will first discuss, in section \ref{LB}, the implementation of the 
LB method to simulate diffusive transport.
We shall propose a new boundary condition which
improves the accuracy and validity of the method,
particularly when describing narrow fractures. 
In section \ref{comparison}
we will compare the proposed boundary condition
with previous ones, in two different cases.
Diffusion (section \ref{diffusion2D}) and Taylor dispersion 
(section \ref{taylor2D})
in a two-dimensional straight channel.

After reviewing some basic facts about self-affine surfaces
(section \ref{self-affine}), 
we investigate, in section \ref{diffusion}, how the
complex geometry of fracture surfaces affects 
the diffusive transport (zero mean flow velocity). We also study
the dependence of diffusivity on the size of the gap 
between complementary 
fracture surfaces. We will show that it is
possible to derive an analytic expression,
which accounts for the geometric effects on the diffusivity,
in the limiting case where the roughness associated
with the fracture surface is small compared to the mean aperture.

In section \ref{sec_dispersion} we study 
dispersive transport in two different situations. 
First, when the two surfaces are simply 
displaced normally to the mean fracture plane. 
In this case, we will
show that a description in terms of a parameter
$\Lambda$, related to {\it dynamically connected}
pore space, accounts for a large part of the enhancement due to
the presence of low-velocity zones in the fracture channel. 
Secondly, we show that when there is a lateral shift as well
as the normal displacement, 
an increase in dispersion is obtained.
Finally, we present a novel result, showing that
dispersion is diminished in the presence of convection
at low P\'eclet numbers.
This effect is found when the surfaces are shifted and the
convective transport is weak enough.


\section{Lattice-Boltzmann model with the BGK collision operator}
\label{LB}

Our goal is to study various aspects of transport in
fractures which are sensitive to the fracture roughness.
Since we consider convection and dispersion in a highly irregular 
geometry, the lattice-Boltzmann method 
\cite{mcnamara,rothman,chen98} is particularly
convenient. In this algorithm, fictitious particles move between
neighboring sites on a lattice, with suitable collision rules, and the
boundary of the flow domain is simply a surface of sites
where boundary condition rules should be imposed. 
We use a version of the LB model
first proposed by Qian {\em et. al} \cite{qian},
with a cubic lattice in 3 dimensions and 19 velocities
(D3Q19 in the terminology of \cite{qian}). 
The collision operator is
approximated by a single relaxation parameter $\lambda$, 
the Bhatnagar-Gross-Krook model\cite{BGK}, 
and the local equilibrium distribution given in
\cite{qian} is used. This pseudo-equilibrium distribution locally
preserves mass and momentum values, and is formulated specifically to
recover the Navier-Stokes equation at large length and time scales.
As usual, we define the lattice spacing as the unit of length,
and the time step in the simulations as the unit time. 
In what follows, quantities will be
measured in terms of the lattice-spacing and simulation time-step.
Note that, since we are concerned with incompressible flows, 
we do not need to introduce a dimension of mass.

The model used for miscible fluids is a straightforward extension
of the model restricted to describe simple fluids.
Following the convention of Rothman and Zaleski \cite{rothman}, we will
distinguish between particle types by assuming that the particles
are colored. The pseudo-equilibrium color distribution 
given in \cite{rothman} is used,
and a single but new relaxation parameter $\lambda_D$ (BGK approximation) 
is used in the equation that describes the advection and 
diffusion of color.   
In addition to mass and momentum, color is also
a conserved quantity.
 
The basic variables of the model are the distribution functions
$N_i$, corresponding to the mean occupation number of particles
in the direction $i$ at a given node, and $\Delta_i$,
describing the relative amount of color \cite{flekkoy93}.
Since the evolution of the total population is independent of the 
color of the particles, the hydrodynamic equations for mass and momentum
may be obtained as in the simple fluids case. In other words,
the time evolution of the $N_i$'s is independent of
the $\Delta_i$'s.
These evolution equations are then coupled to the evolution of color, 
through the local fluid velocity.
Thus, the diffusive behavior of the fluids is superimposed on the 
underlying Navier-Stokes dynamics.

The fact that that the information on mass density and flow
is decoupled from the information on color, opens the possibility 
of imposing different boundary conditions (BC) on the populations 
describing fluid flow ($N_i$) and those describing advective-diffusive 
transport ($\Delta_i$). 
Specifically, for the non-slip BC on the solid surface,
we shall use the simplest implementation of particle exclusion -
the bounce-back rule (BB), where the particles incident on the boundary are 
propagated back into the directions from which they came.
On the other hand, for color concentration a 
different macroscopic BC is desired,
{\em i. e.} zero color-gradient normal to the solid surface.

\subsection{Diffusion boundary conditions}
\label{BC-D}

As mentioned, we will study tracer dispersion in narrow
gaps between self-affine solid surfaces. Thus, in this situation,
the BC imposed at the solid surfaces becomes a crucial aspect 
of the simulation method. 

In previous works concerning lattice BGK models for miscible fluids,
BC imposed on color concentration at solid boundary 
were not distinguished from those used to simulate fluid flow.
In these situations, the relevance of the BC at solid surfaces  
varies depending on the particular system under consideration.
In some cases, when simulating bulk processes, as in \cite{flekkoy93,holme92},
there is no need to treat the BC separately, and periodic boundary conditions
on all distribution functions may be used. 
In other cases, fully three dimensional problems in Helle-Shaw cells
are described in terms of two dimensional LB model by
the modification of the forcing to account for the viscous drag 
of the top and bottom plates of the cell \cite{oxaal94,flekkoy95,flekkoy96}. 
Therefore, molecular diffusion
in the transverse direction is not described and similar BC
can be applied to all fields.  
In the others, where fluid flows in complex geometries \cite{gutfraind95} or
in narrow channels bounded by solid surfaces \cite{baudet89,cali92}, the 
above-mentioned distinction in BC would be desirable.

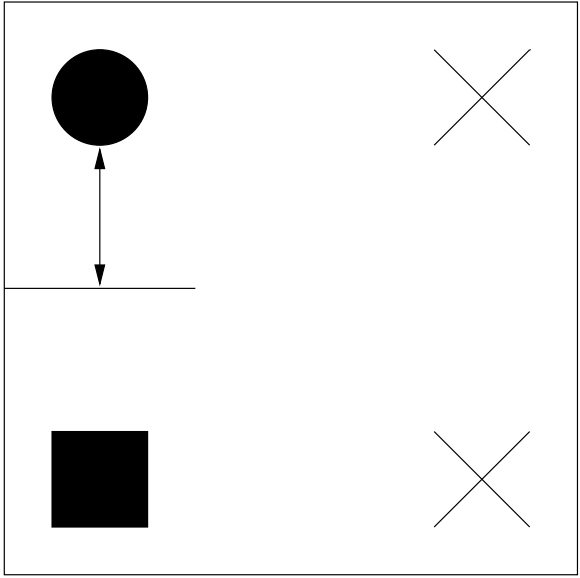
\begin{figure}
\caption[bf]{Schematic view of the bounce-forward collision rule.
Solid squares represent solid sites, open circles represent 
fluid sites, and solid circles represent particles. Solid lines 
correspond to fluid-solid interface, situated half-way 
between solid and fluid sites. Arrow lines represent 
incoming and outcoming trajectories.}
\label{bf}
\end{figure}

Here we will use a different set of BC for color concentration, 
to ensure zero color-gradient normal to the solid surface.
To this end, we implement a mirror-reflection condition or
bounce-forward rule (BF), where upon collision only the normal
component of the particle velocity is changed.
In Figure~\ref{bf} we show a schematic 
view of the BF collision rule. The last case was chosen
arbitrarily, due to its very simple implementation, among 
the different possibilities that satisfy the 
mirror-reflection-type condition.
For instance, it would be also consistent to split the 
incoming concentration into the two 
neighboring sites.
  
Physically, BB has the (approximate) effect of making both components
of the fluid velocity vanish at the solid, consistent with
the non-slip condition. A passive tracer may, however, diffuse along
a solid boundary, so the relevant condition
is that only the normal component of the flux vanishes
at a solid. the bounce-forward rule is the obvious LB realization of
the physical boundary condition.

To answer the question whether this BC 
improves the LB method for miscible fluids we shall present the results
of numerical simulations using both BB and BF in two
different situations: diffusion and Taylor hydrodynamic dispersion
in a two-dimensional straight channel.

\section{Comparison between BB and BF rules: numerical simulations}
\label{comparison}

We will exhibit a comparison between BB and BF,  
in two applications, diffusion and Taylor dispersion in 
a two-dimensional channel. These situations are particularly 
important in terms of addressing the validity and accuracy
of the above-mentioned boundary conditions, particularly 
if one is interested in further use of them in complex geometries.

\subsection{Diffusion in a two-dimensional straight channel}
\label{diffusion2D}

Diffusion was studied varying the channel width $H$ 
from $4$ to $40$ in grid sites, in a channel of length 
$L=512$, with periodic BC used at both ends.
In the LB model the diffusion coefficient in bulk is given by,
\begin{equation}
D_m = \frac{1}{3} \left( \frac{1}{\lambda_D} - \frac{1}{2}  \right)
\end{equation}
varying the relaxation parameter $\lambda_D$ we 
have also studied diffusive transport for a set of bulk diffusivities
ranging from $D_m=8.8 \times 10^{-4}$ to $D_m=1.5$.

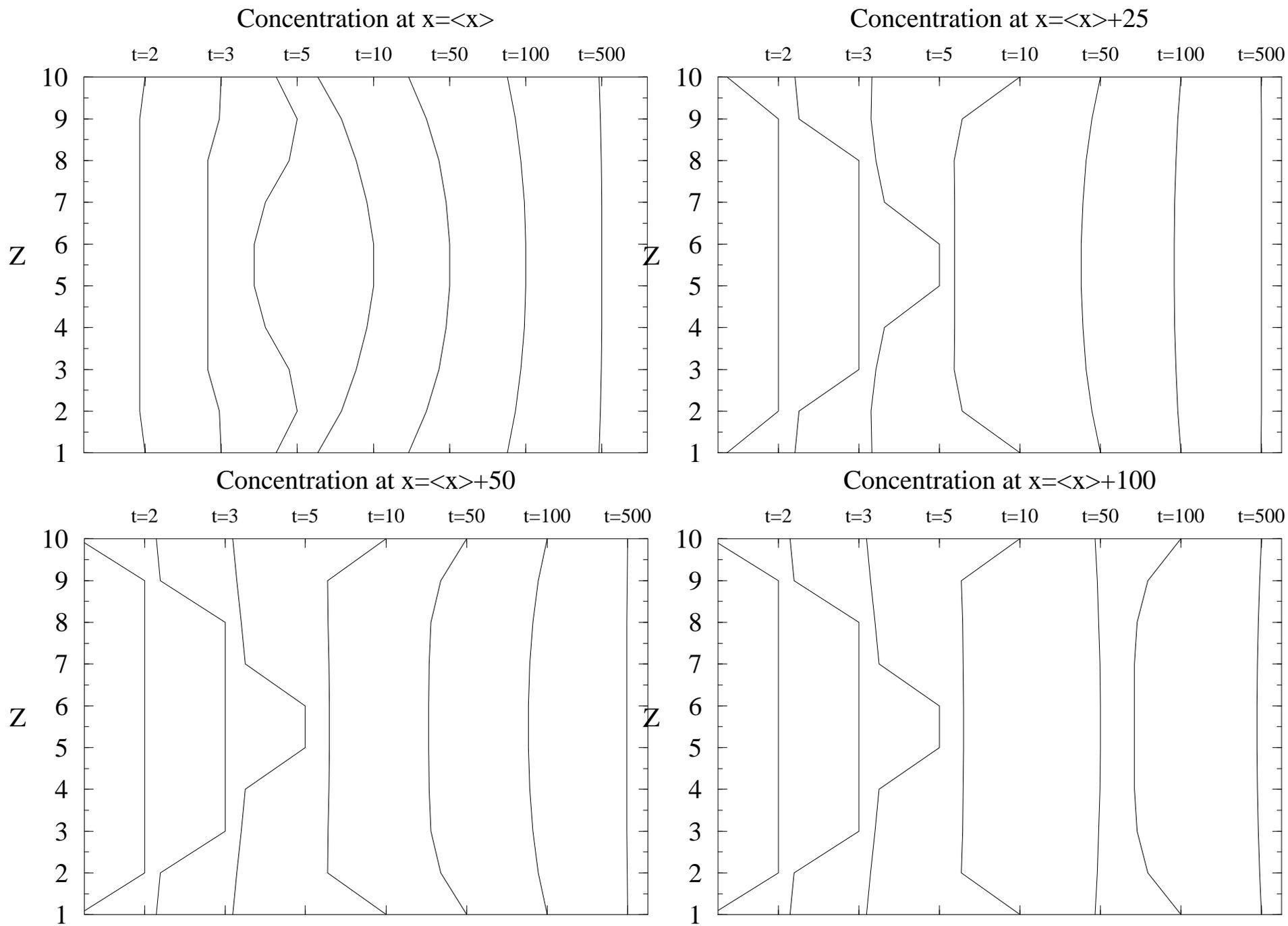
\begin{figure}
\caption[vertical]{
Vertical dependence of the color-concentration. Horizontal axes 
represent normalized concentration. For clarity the curves corresponding to 
different times are shifted horizontally by a constant value. 
Results correspond to numerical simulations in a channel of 
width $H=10$ and length $L=512$.
The initial condition is a Gaussian color-distribution 
centered at $x=256$ and with $\sigma^2=4.0$.
The relaxation parameter is $\lambda_D=0.2$ corresponding
to a bulk diffusion coefficient $D_m=1.5$.}
\label{vertical}
\end{figure}

In all the simulations for both boundary rules,
the obtained depth-averaged color-concentration 
displays a Gaussian distribution,
and the mean square displacement grows linearly in time.
However, the vertical dependence of the concentration, 
as well as the diffusivity values, 
strongly depend on the BC used.
While concentration remains vertically-homogeneous
using BF, in Fig.~\ref{vertical} we show that,
using BB yields the undesired effect of a transient
vertical variation of color.
Two different situations can be distinguished
in Figure~\ref{vertical}, namely, 
close to the mean position of tracer particles $\langle x \rangle$, 
or far from it. (Note that $\langle x \rangle$ is constant in time).
At $\langle x \rangle$ (Fig.~\ref{vertical}a), the initial 
concentration gradient in drives tracer particles. 
Due to the bounce back rule, particles close to the surface
remain there longer, and the relative tracer concentration
builds up. Shortly after, particles
close to the surface diffuse toward the center of the channel,
and eventually the concentration gradient in vertical direction
is smoothed out. On the other hand, far from the mean
position of tracer particles the situation is inverted.
Bounce back at solid surfaces, makes the tracer particles mostly arrive
from the center of the channel, and a vertical concentration
gradient develops. Thus, after several time steps, particles
diffuse from the center toward the solid surfaces. At large
times, the vertical gradient vanishes an concentration
becomes homogeneous across the channel. 

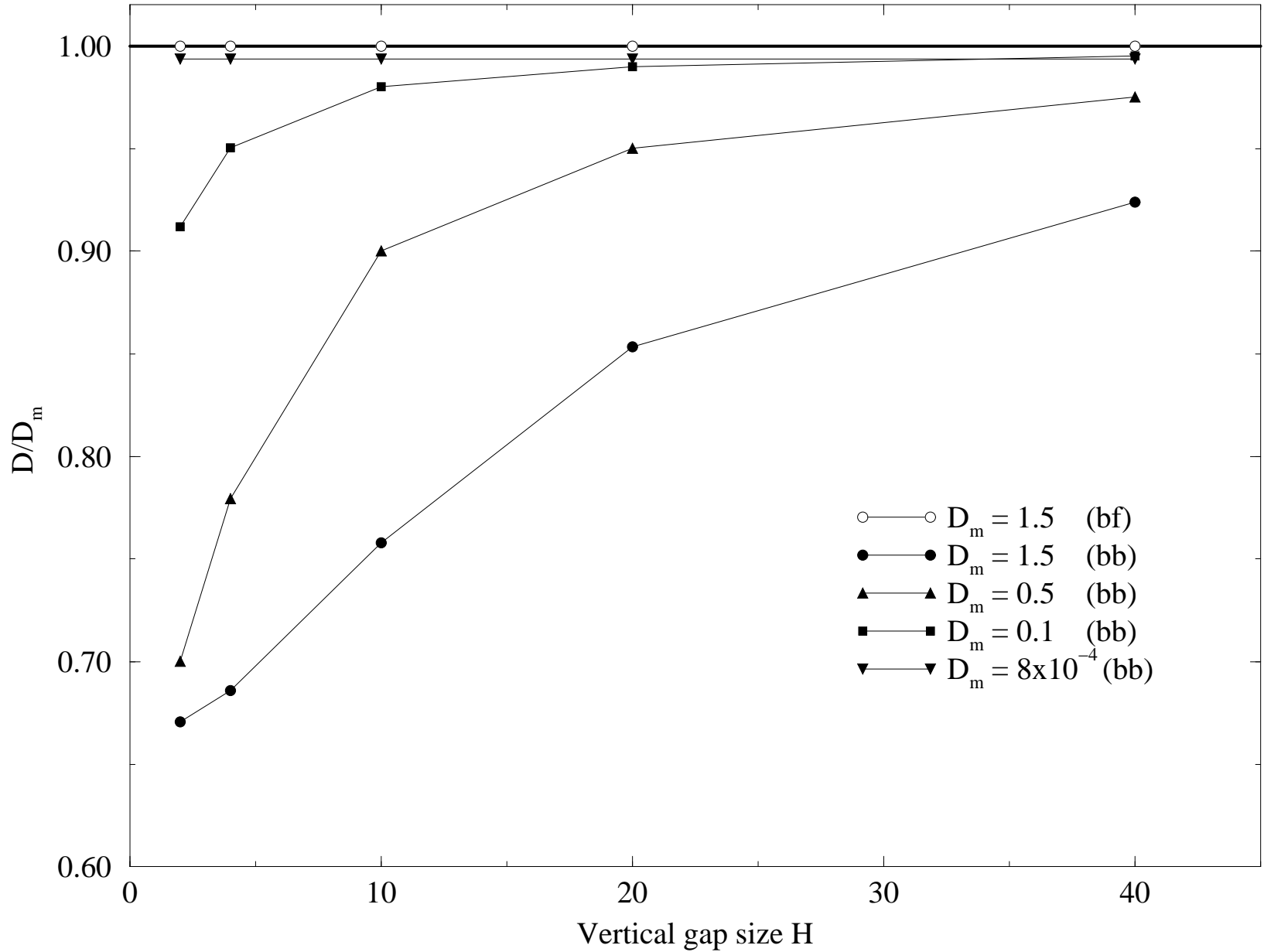
\begin{figure}
\caption[dreal]{ Normalized diffusion coefficient $D/D_m$ as a 
function of the vertical gap size. Filled symbols correspond to simulations 
using BB (for different bulk diffusivities). Open circles correspond 
to simulations using BF and $\lambda_D=0.2$. Results were obtained 
in a straight channel of length $L=512$. The initial condition is 
a Gaussian color-distribution centered at $x=256$ 
and a width $\sigma^2=4.0$.}
\label{d_real}
\end{figure}

Using BB, the diffusion coefficient 
strongly depends on the vertical size of the channel.
Diffusive transport close to the solid surfaces is reduced due to the 
BB rule, yielding a smaller diffusion coefficient compared to
bulk diffusivity. This effect becomes negligible when the gap is large enough
or, when the diffusivity is sufficiently small.
In Figure~\ref{d_real} we show the dependence of the 
diffusion coefficient (relative to $D_m$), 
on the size of the gap and bulk diffusivity.
We also show that simulations using BF give $D=D_m$, 
for any size and $D_m$.

As mentioned, the undesired effects due to BB boundary conditions
become less significant at small diffusivities. When using
$D_m=8.8 \times 10^{-4}$ the deviation of the numerical value
from the expected one is within $1\%$. However, the use of
small diffusion coefficients allows large concentration 
gradients, which may result in numerical instability.
Discrepancies between theory and simulations have been reported, 
in LB model of miscible fluids, for values of $D_m$ 
below $10^{-4}$\cite{flekkoy93}. Also, for values of $D_m$ below $10^{-4}$,
oscillations between negative and positive values of concentration have been
observed \cite{flekkoy95}.
Therefore, the parameter space where 
the model is fairly independent of the particular choice 
of the boundary condition rule, is very close to being
numerically unstable. 
 
\subsection{Taylor dispersion}
\label{taylor2D}

Finally, we compute the asymptotic hydrodynamic dispersion 
when a mean flow is set within the channel. 
In this case, dispersion has two different contributions, 
one due to molecular diffusion and the
other due to spatial variations of the fluid velocity, namely
Taylor dispersion \cite{taylor}. 
Asymptotic analysis of Taylor dispersion in
a two-dimensional channel of constant width $H$ and 
mean flow velocity $U$
gives the exact formula for the longitudinal dispersion
coefficient\cite{aris59},
\begin{equation}
\label{taylor_dispersion}
D_{\parallel} = D_m + \frac{1}{210} \frac{U^2 H^2}{D_m},
\end{equation}

Introducing the P\'eclet number, which accounts for the relative 
magnitude of the convective and diffusive transport,
$\rm{Pe}=U H / D_m$, the previous equation can be rewritten as,
\begin{equation}
\label{peclet}
D_{\parallel} = D_m \left( 1 + \frac{\rm{Pe}^2}{210} \right).
\end{equation}

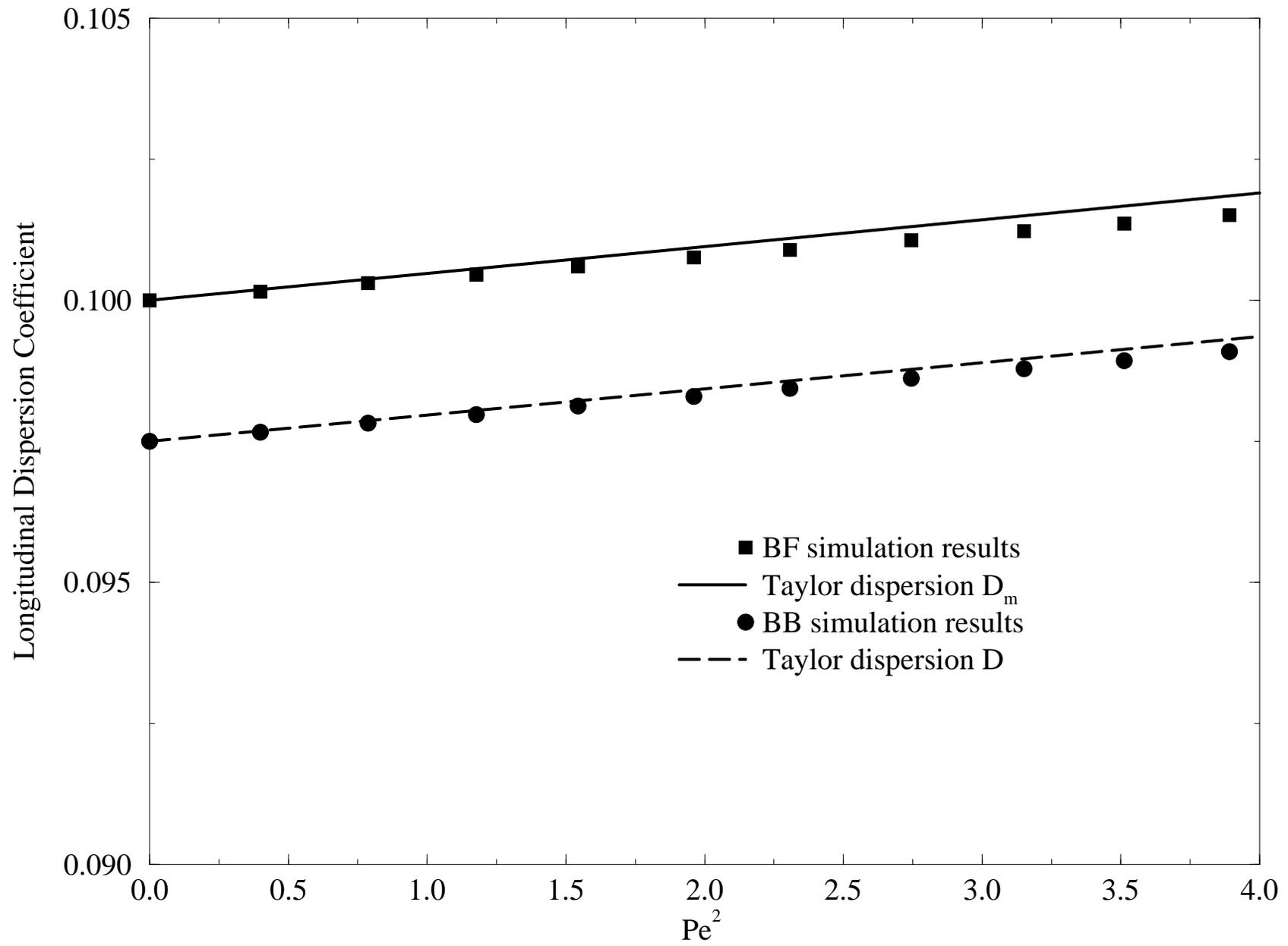
\begin{figure}
\caption[taylor]{Longitudinal dispersion coefficient as a function of
the P\'eclet number squared. Solid line corresponds to Eq. (\ref{taylor_dispersion}). 
Dashed line also corresponds to Eq. (\ref{taylor_dispersion}) but 
using $D$ measured in diffusion simulations (instead of $D_m$). 
Solid squares and circles refer to the numerical
results using BF and BB respectively. The channel grid is
$16 \times 1024$ and the diffusion coefficient in bulk is $D_m=0.1$.}
\label{taylor_fig}
\end{figure}

Concerning the numerical simulations, whether the BF rule is implemented or
not, the Taylor regime for the longitudinal dispersion is expected to hold.
However, the effective diffusivity is slowed down when  
BB collision at solid surfaces is used. Thus, as the diffusivity is
smaller, the characteristic homogenization time 
in the transverse direction 
becomes larger. In other words, we may expect Eq. (\ref{taylor_dispersion})
to hold, but where $D_m$ is replaced by the effective diffusion coefficient
in the channel $D$, as measured in diffusion simulations using BB rule.
(section \ref{diffusion2D}).
In Figure ~\ref{taylor_fig} we compare the numerical results with
Eq. (\ref{taylor_dispersion}).
There is a very good agreement between the numerical results and 
the Taylor theory in both cases, using BB and BF as boundary rules.
Let us remark that solid and dashed lines corresponds exactly to 
Eq. (\ref{taylor_dispersion}) with no adjustable parameter,
and the only difference between them is the molecular diffusivity value.
These results confirms the previous one, showing that the BB rule
diminishes the diffusive transport and consequently enlarges 
the longitudinal hydrodynamic dispersion. A departure from Eq. 
(\ref{taylor_dispersion}) at large P\'eclet numbers can also be observed, 
where the largest discrepancy is $0.3\%$. Similar 
results, showing slightly smaller numerical values 
than theoretical ones have been previously reported \cite{cali92}. 

As a conclusion to this section we may say that the effect of using
the same BC on fluid and color distributions slows
down the diffusive transport. This effect becomes 
appreciable when the system is narrow enough or the diffusion 
coefficient is large.

\section{Self-affine narrow fractures}
\label{self-affine}

We briefly review here the mathematical characterization of self-affinity.
A more detailed discussion can be found in our previous work
\cite{GK1}.
We consider a rock surface without overhangs, whose height is given by a
single-valued function $z(x,y)$, where the coordinates $x$ and $y$ lie
in the mean plane of the fracture.  Self-affine surfaces \cite{feder} 
display scale invariance with different dilation ratios along different 
spatial directions (in contrast to self-similar surfaces 
which stretches the coordinates with equal ratios). 
Experiment indicates that isotropy can be assumed in 
the mean plane, implying that there
is only one non-trivial exponent relating the dilation ratio
in the mean plane to the scaling in the perpendicular direction, {\em i.e.},
\begin{equation}
\label{surf_scaling}
z(x,y) = \lambda^{-\zeta} z(\lambda x, \lambda y)
\end{equation}
where $\zeta$ is the roughness or Hurst exponent \cite{mandelbrot}. 
In all cases the roughness exponent is chosen as the 
experimentally-observed value $\zeta=0.8$.

We shall emphasize the 
limiting situation of narrow fractures, 
in which the two surfaces are very close to each other. 
Consider the situation in which a rock of lateral size $L$ is fractured
and the two surfaces are simply displaced by a distance $H \ll L$,
perpendicular to the mean fracture plane, with no relative shift.
The fluctuations (the difference between the maximum and 
minimum values of $z$) scales as $R\sim L^\zeta$. If $H$ is the width of the fracture, 
then the limiting situation corresponds to $R \gg H$.

In narrow fractures, the correlation between the two sides is 
an important feature for the dispersion process, and we
consider two possibilities.  We first study fractures where the upper 
surface has been simply translated a distance $H$ normal to the mean plane,
so that the local aperture $a(x,y)$ equals the constant $H$. 
Alternatively, the two surfaces may have a relative lateral displacement $d$
in the mean fracture plane, accompanied by a displacement $H$ in the 
perpendicular direction, so that the two surfaces do not overlap.
In this case the local aperture is given by the random variable.
\begin{equation}
\label{aperture}
a_d(x,y) = z(x+d,y) - z(x,y) + h
\end{equation}
It turns out \cite{plouraboue95} that $d$ is the lateral correlation length 
for fluctuations in the aperture, in the sense that $a_d(x,y)$ and 
$a_d(x+\Delta x,y)$ decorrelate for $\Delta x\gg d$.

The notion of fractal dimension also differs from the 
case of self-similar surfaces. 
It has been shown \cite{mandelbrot86I,mandelbrot86II,mandelbrot86III},
that one needs in general several distinct notions.
We are interested in the local and
global box dimensions of vertical plane sections.
Using the self-affine scaling law for the correlation function,
\begin{equation}
\label{variance}
\sigma_z^2(\Delta x) =
\langle [z(x,y)-z(x+\Delta x,y)]^2 \rangle = \phi(\ell) 
\left({\Delta x \over\ell} \right)^{2\zeta}
\end{equation}
where $\ell$ is a microscopic length, say a grain size, such that
\begin{equation}
\label{grain_size}
\sigma_z^2(\ell) = \phi(\ell) \sim \ell^2,
\end{equation}
we can estimate,
\begin{equation}
\label{Dz}
\Delta Z(\Delta x) \equiv z(x,y)-z(x+\Delta x,y) 
\sim \ell \left({ \Delta x \over \ell} \right)^\zeta.
\end{equation}

For length scales $\Delta x \ll \ell$ it can be seen from the 
previous relation that $\Delta Z \gg \Delta x$ (given that $\zeta$
is smaller than one). This so called local property
yields a local box counting dimension $D_{BL}=2-\zeta$. 
On the other hand, for length scales $\Delta x \gg \ell$,
it is clear that $\Delta Z \ll \Delta x$. Thus,
the global box counting dimension is $D_{BG}=1$. 
As we consider $\ell$ as the lower cut-off of the self-affine
behavior in real fractured system, we will focus on the
global regime where the box dimension is one. This fact
largely determines the convective-diffusive behavior, 
as we will discuss later.

In this paper we restrict ourselves to the two-dimensional case 
where the surface is invariant in the $y$ direction, $z(x,y)=z(x)$, 
and the mean flow, when present, is forced in the 
$x$ direction by a constant pressure drop. 
In a subsequent paper we will extend these calculations to fully three
dimensional fractures, but it is convenient, both conceptually and in the
numerical simulations, to regard the system as having a translationally
invariant third dimension.  

We use statistically self-affine surfaces
with periodic boundary conditions.  The periodicity is not a physically
essential ingredient here, but has some calculational advantages in 
alleviating finite-size effects.  The surface is generated by a Fourier 
synthesis method, based on power-law filtering of arrays of 
independent random numbers \cite{GK1,voss85}. 

\section{Diffusion in narrow fractures}
\label{diffusion}

In this section we study diffusion in two-dimensional
self-affine fractures. The approach will be based on the
tortuosity concept \cite{bear}. It is a common characteristic
in fluid transport in porous media, and in fractures in particular,
that the actual path followed by the fluid is very tortuous.
In our fractured two-dimensional systems, 
a purely geometrical definition of tortuosity can be
done, by considering the ratio of the shortest
continuous paths between any two points within the fracture to
the length of the system projected on to the main plane
\cite{adler}.
It is clear that, for narrow fractures, this ratio
will approach the ratio between the length of the
surface profile in the $xz$ plane to the distance 
$\Delta x$ between the two points.
Then, if $l_e$ is the true length between the two points
separated a distance $l=\Delta x$, we can write for the tortuosity,
\begin{equation}
\label{tortuosity}
T = \frac{l_e}{l}  
\end{equation}

The tortuosity factor $T$ accounts for the reduction in diffusivity, 
due to irregular geometries, in a very simple way.
Along the actual average-path of tracer particles, diffusion behaves
as in free bulk, where the diffusion coefficient is the free bulk
molecular diffusion coefficient $D_m$. 
On the other hand, diffusive spreading along the $x$ axis
is reduced, due to the difference between $l_e$ and $l$.
In terms of $T$, this difference can be accounted for by writing,
\begin{equation}
\label{effective_diffusion}
D = \frac{l^2}{2 t} = \left( \frac{l}{l_e} \right)^2 \frac{l_e^2}{2 t} = 
\frac{D_m}{T^2}
\end{equation}
(some authors prefer to define tortuosity as $T=(l_e/l)^2$, or
as the inverse of this definition \cite{bear,dullien}).

For the previous equation to be meaningful one should have
a constant value of $T$, that is, independent of $l$.
Whether or not $T$ is constant clearly depends, for
narrow fractures, on the dimension
of the vertical plane sections of the surface. 
In general, the effective length $l_e$ depends on $l$ through
$l_e \propto l^{D_{B}}$, where $D_B$ is the dimension
of the surface profile. Therefore, while at local
length scales $D_B=D_{BL}=2-\zeta$ and $T$ depends on $l$,
at global length scales, considered in this work,
$D_B=D_{BG}=1$ and the tortuosity factor is constant
\cite{mandelbrot86I,mandelbrot86II,mandelbrot86III}.
For the numerically-generated self-affine surfaces used in this work,
we tested the dependence of the effective length $l_e$ on $l$.
Using the method of `dividers' \cite{mandelbrot} to measure $l_e$, we obtain, 
in all cases, a linear relation between the dividers
opening $\epsilon$ and $l_e(\epsilon)$, {\em i.e.} $D_B=1$. 

The fact that $T$ is constant has some 
important consequences on diffusive transport.
First, the relation between mean square displacement and
time should be linear (after perhaps a transient time).
Second, the distribution of tracer concentration
should be asymptotically Gaussian, 
and independent of the initial distribution of tracer.
Finally, the correction due to tortuosity is strictly
geometrical and therefore it should be independent on
the actual value of the free bulk molecular diffusivity $D_m$.
 
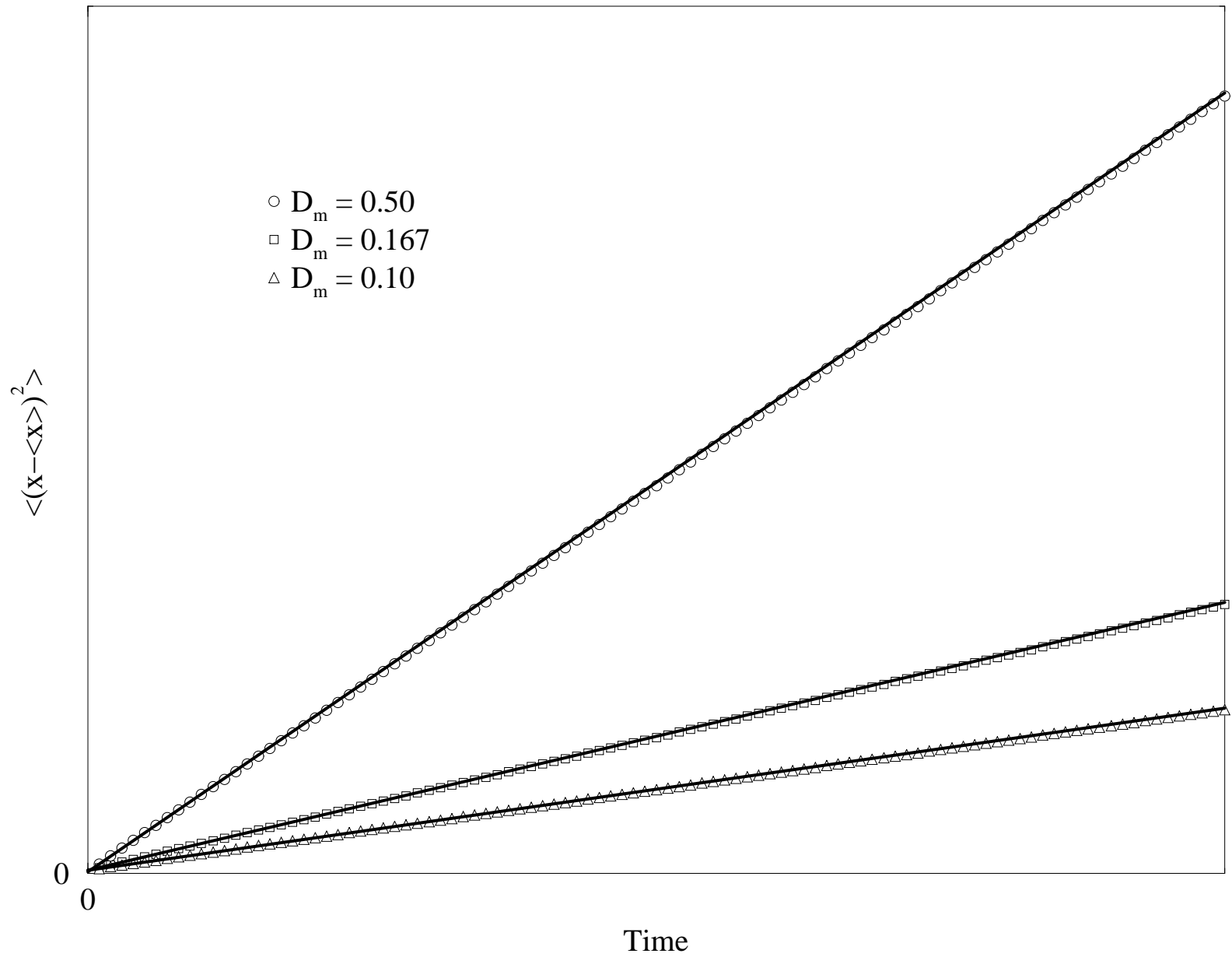
\begin{figure}
\caption[linear]{Mean square displacement as a function of time for
different diffusion coefficients. Simulations where performed in a
system of length $L=512$ and separation between surfaces $H=4$.}
\label{linear}
\end{figure}

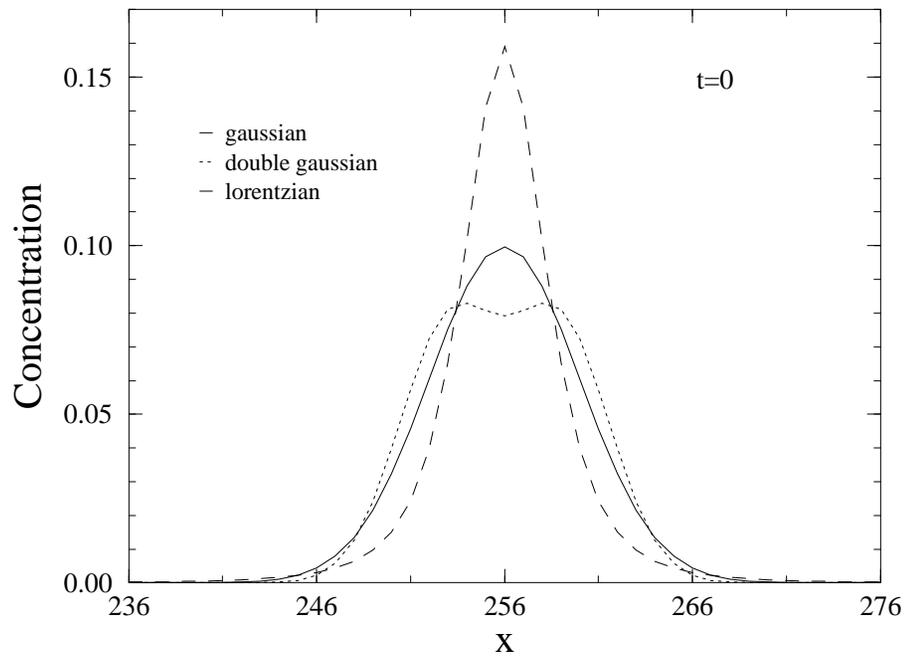
\begin{figure}
\caption[gaussian]{Evolution of tracer concentration for different initial
conditions: a Gaussian distribution, two Gaussians centered at
$x_0 = 256 \pm 3$, and a Lorentzian distribution squared.
The initial dispersion is in all cases $\sigma_0=4.0$ and $D_m=0.5$.}
\label{gaussian}
\end{figure}

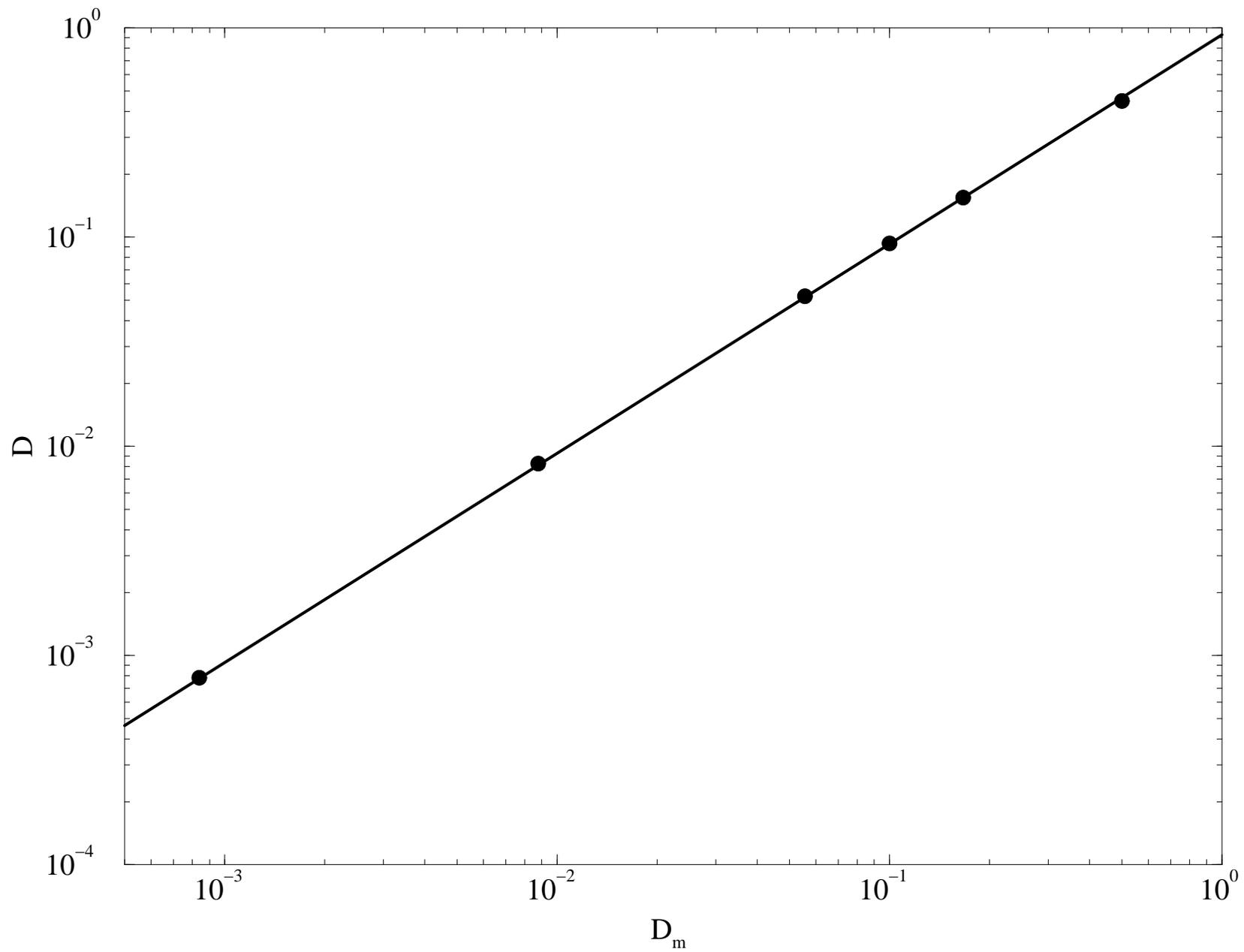
\begin{figure}
\caption[tortuosity]{Diffusion coefficient in the fractures $D$ 
as a function of the free bulk diffusion coefficient $D_m$. The system
length and width are $L=512$ and $H=16$. Solid circles
correspond to simulations with $D_m = 5.0 \times 10^{-1} ; 
1.67 \times 10^{-2} ; 1.0 \times 10^{-1} ; 8.77 \times 10^{-3}$ and 
$8.38 \times 10^{-4}$. The solid line
corresponds to Eq.~(\ref{effective_diffusion}) with
a tortuosity $T^2=0.93$.}
\label{f_tortuosity}
\end{figure}

In Fig.~\ref{linear} we show the linear dependence of the mean 
square displacement on time, for different values of the diffusion coefficient. 
Many other simulations, with different values of $D_m$ and $H$ were performed,
and in all cases a linear relation was obtained. In Fig.~\ref{gaussian}
we show how the same Gaussian distribution is approached at long times, 
starting from three different initial distribution of tracer particles.
Finally, in Fig.~\ref{f_tortuosity} we show that the tortuosity factor
is a strictly geometrical property, being clearly independent on 
the diffusivity of the tracer particles.

\subsection{Tortuosity dependence on the fracture width}
\label{tortuosity_vs_H}

In the previous section, we discussed the linear relation between 
the effective length $L_e$ and $L$, and its consequences on 
diffusive transport. We showed that the effect of surface roughness on
diffusion can be accounted for by a purely geometrical property
of the system, the tortuosity $T$.
However, it remains to analyze the dependence of $T$ on 
geometrical parameters that describe the system. 
Of particular importance, is the dependence of $T$ on the 
width of the fracture $H$.

The theoretical analysis presented here will closely follow 
the kind of approximation used in \cite{GK1}, where the fracture 
is divided into a sequence of quasi-linear segments at 
varying orientation angles. First, we estimate
a typical size $\xi_{\parallel}$ in the direction of the mean flow over which
the fluctuations in the vertical direction
are small compared to the effective aperture of the channel.
A segment of length $\xi_\parallel$ is roughly straight when
$\Delta Z(\xi_{\parallel}) \sim \ell (\xi_\parallel/\ell)^\zeta$
is a small fraction of $H$, which yields
\begin{equation}
\label{delta-h}
\xi_\parallel \sim \ell \left({ H \over\ell }\right)^{1/\zeta}
\end{equation}
 
Returning to the entire fracture, 
each $\xi_{\parallel}$-channel is oriented at some angle $\theta_i$ 
with respect to the mean plane, and has effective aperture 
$a_i=H \cos\theta_i$, and length 
$\xi_{\parallel}^i=\xi_{\parallel}/\cos\theta_i$. 
Thus, we write for the total length of the channel
\begin{equation}
\label{le}
L_e = \sum_{i=1}^N \xi_{\parallel}^i 
= \xi_{\parallel} \sum_{i=1}^N \cos^{-1}(\theta_i)
\end{equation}
Finally, taking into account that $N=L/\xi_{\parallel} \gg 1$ is 
the number of channels,
we can convert the sum into an average over the distribution of angles, 
and write for the tortuosity,
\begin{equation}
\label{T}
T = \frac{L_e}{L} =  
{\langle \cos^{-1}(\theta) \rangle}
\end{equation}

Since $\zeta<1$, the channels have small vertical fluctuations, and we 
can give a simple estimate of the cosine as
\begin{equation}
\frac{1}{\cos\theta} = 
\frac{\sqrt{\xi_{\parallel}^2 +\Delta Z^2(\xi_{\parallel})}}{\xi_{\parallel}}
\approx 
1+ \frac{1}{2} \left(\frac{\Delta Z(\xi_\parallel)}{\xi_\parallel}\right)^2
\end{equation}
Replacing this approximation in Eq.~(\ref{T}), we obtain,
\begin{equation}
\label{T_narrow}
T \approx \left[ 1 + \frac{1}{2}
\left(\frac{\sigma_z(\xi_\parallel)}{\xi_\parallel}\right)^2 \right]
\end{equation}
where $\sigma^2_z(\xi_\parallel)=
\langle \Delta Z^2(\xi_\parallel) \rangle$ 
(see Eqs.~(\ref{variance}) and (\ref{Dz})).

A more precise evaluation can be obtained 
based on a Gaussian distribution of 
heights \cite{GK1}, as supported by experimental
measurements \cite{plouraboue95} and, in fact, 
the actual distribution given by
our numerical procedure for generating self-affine surfaces.
In this case, the angular average is given by
\begin{equation}
\label{cosine}
\langle \cos^{-1}\theta \rangle = 
\int p(\Delta Z) \sqrt{1+\left( \frac{\Delta Z}{\xi_\parallel} \right)^2}
= x^{1/2} U\left(\frac{1}{2};2;x\right)
\end{equation}
where $x=(\xi_\parallel^2/2\sigma_z^2(\xi_\parallel))$, and
$U(a;b;x)$ is the {\it confluent hypergeometric
function of the second kind}\cite{lebedev,abramowitz}. 
Therefore, we can get a more convincing evaluation of 
the tortuosity of a narrow two-dimensional self-affine fracture, 
using the leading terms in the
asymptotic representation of $U$ for large 
$x$ and Eq.~(\ref{delta-h}) for the value of
$\xi_\parallel$,
\begin{equation}
\label{T_final}
T = \left[ 1 + \left( \frac{\phi(\ell)}{\ell^2} \right)
\left( \frac{H}{\ell} \right)^{\frac{2\zeta-2}{\zeta}} \right] 
\end{equation}

To compare the previous relation with the numerical results, let us first
recast it in terms of the diffusion coefficient,
\begin{equation}
\label{Dteo}
\frac{D_m-D}{D_m} \approx 2~C_1 \left( \frac{H}{\ell} \right)^
{\frac{2\zeta-2}{\zeta}}
\end{equation}
where we have added an adjustable parameter $C_1$, which is expected to 
be of order one. 

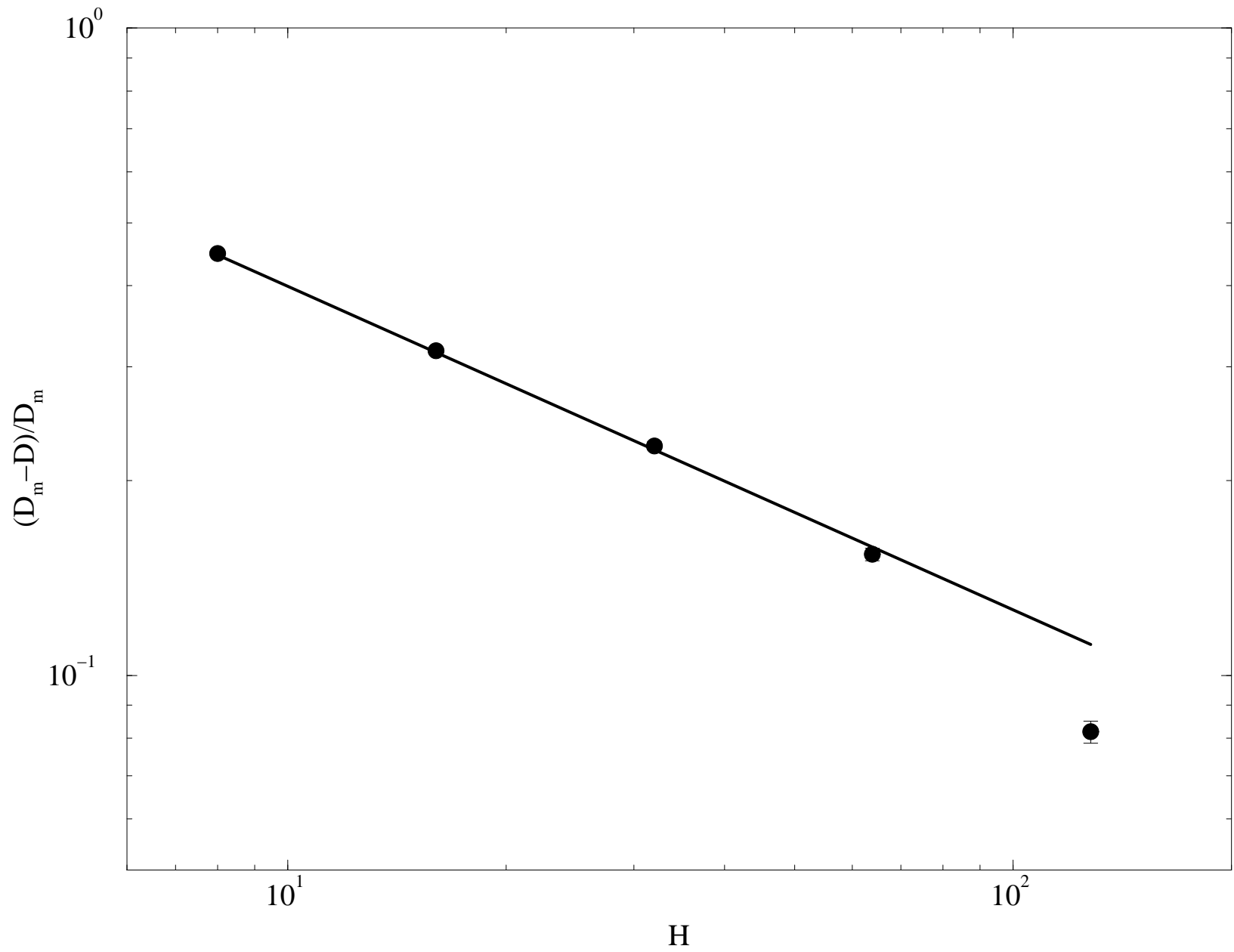
\begin{figure}
\caption[DH]{Correction to the diffusivity due to tortuosity 
of the fractures as a function of the gap $H$.The solid line correspond to 
Eq.(\ref{Dteo}) with $C_1=0.9$. The simulation parameters are
$L=512$, $\lambda_D=0.5$ and $\zeta=0.8$}
\label{DvsH}
\end{figure}

In Fig.~\ref{DvsH} we present the decrease in the diffusivity due to the
tortuosity of the channel as a function of the distance between
the opposite surfaces $H$. We find a good agreement for the predicted
exponent $(2\zeta-2)/\zeta=-1/2$ (the roughness exponent is $\zeta=0.8$).
The only adjustable parameter is the coefficient $C_1$,
which is found to be $C_1 \approx 0.9$ in good agreement
with the expected value.

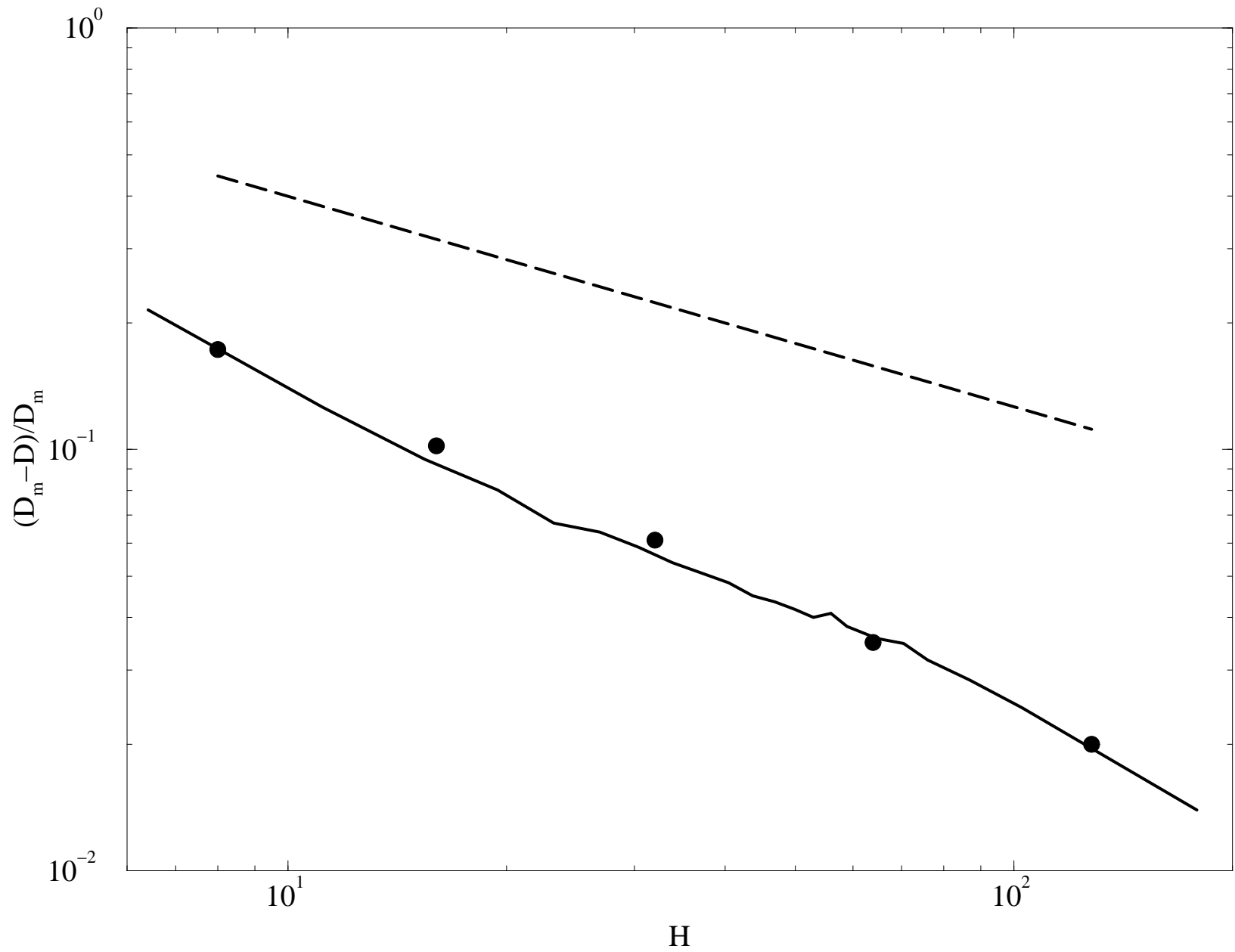
\begin{figure}
\caption[S]{Correction to the diffusive transport when the 
surface fluctuations are small. The solid line is the correction due to 
the tortuosity of the channel, measured from the numerically-generated surfaces.
The dashed line correspond to the observed reduction in diffusivity when
the fluctuations are larger and the dependence is correctly reproduced
by Eq.~(\ref{Dteo}).}
\label{small_s}
\end{figure}

In Fig.~\ref{small_s} we present the numerical results 
obtained for smaller fluctuations in the surface height.
In this case it is clear that the exponent differs from the
predicted value. We believe that, this 
discrepancy comes from the discretization of the surface
height which introduces a larger error the smaller
the fluctuations are. However, we also show in Fig.~\ref{small_s},
the theoretical correction to diffusive transport
computing the tortuosity factor directly through the
numerically-generated surfaces instead of using the asymptotic 
analysis. In this case a good agreement is recovered and
the only adjustable parameter $C_1\approx 2$ is again of order
one (and in agreement with previous results \cite{GK1}).
Similar results were obtained using several 
intermediate values for the surface fluctuations 
in height, where the reduction in diffusivity is always in
agreement with the correction due to the tortuosity
of the channels. In addition, we found that the decrease 
in diffusivity is, in all cases, 
well described by a power-law (where the
power-law exponent increases slightly with the amplitude of the
surface fluctuations), for which we have no explanation.


\section{Dispersion in narrow fractures}
\label{sec_dispersion}

In this section we address the case of Taylor
hydrodynamic dispersion in narrow fractures.
First, we will analyze the case where the two sides
of the fracture are displaced normally to the mean fracture
plane, and then we will turn to the case where there is
a lateral shift as well.

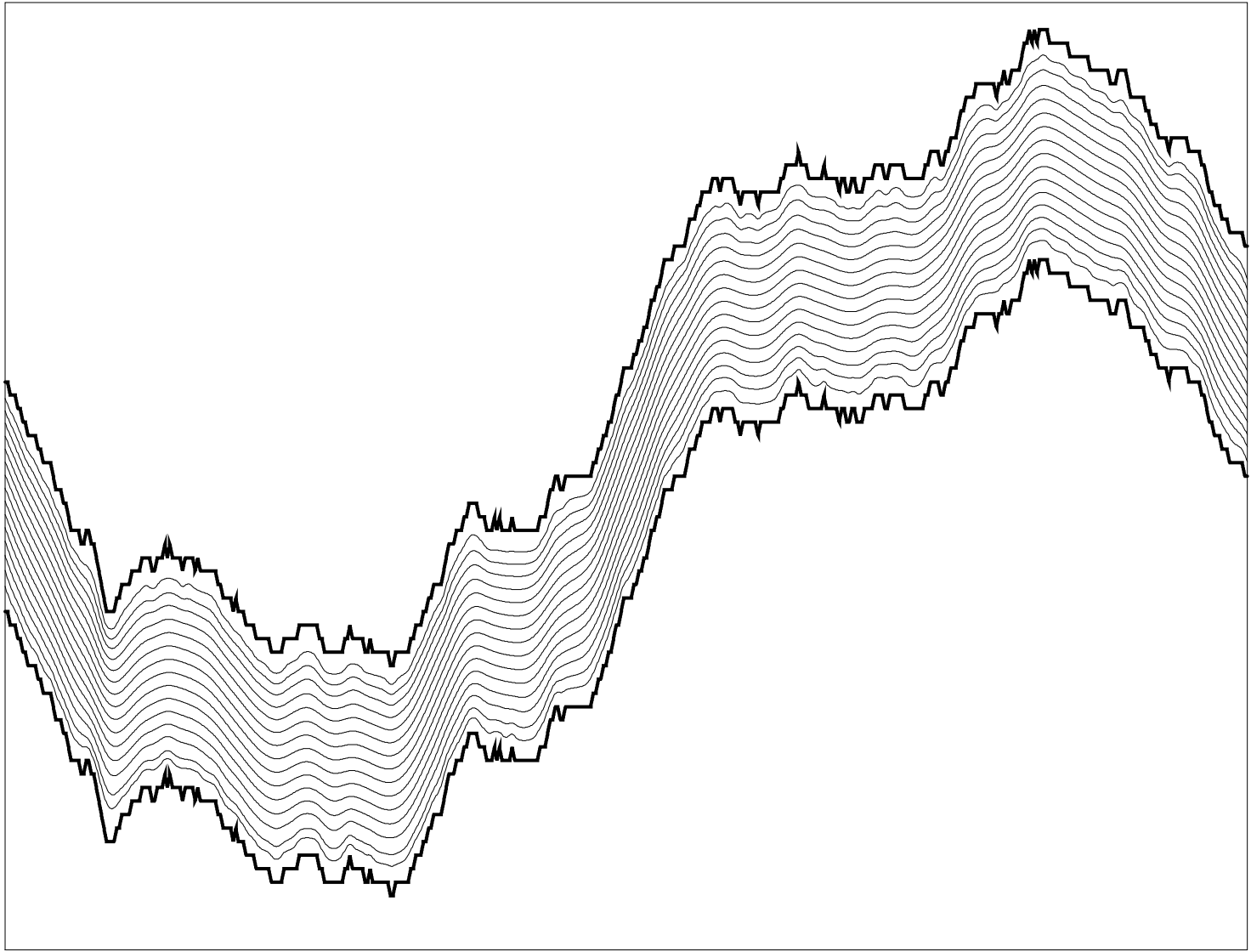
\begin{figure}
\caption[stream]{Streamlines inside a fracture of width $H=16$, length 
$L=512$ and no lateral shift between complementary surfaces. 
Note that, the closer the angle of the surface to $\pi/2$ the narrower 
the effective width becomes.}
\label{stream}
\end{figure}

When the two complementary surfaces are simply displaced vertically by
a distance $H \ll L$, the vertical aperture of the system is constant
everywhere. Nevertheless, the flow field differs from one in a straight
channel, due to variations in the local width of the channel
normal to the mean flow direction. In Fig. \ref{stream}
we show a set of streamlines inside a fracture, where the effect
of the varying effective aperture of the fracture is evident 
(aperture normal to the mean flow).
Moreover, in \cite{GK1} we show that the complex geometry of the fracture gives
rise to low-velocity zones (close to depressions and corners),
reducing the permeability of the system.
In order to describe the dispersion process we need to 
obtain a measure of the fraction of the system that is
subject to convection. To this end we will make use
of the $\Lambda$ parameter \cite{johnson86,johnson87}, 
which is directly related to
transport, and measures the {\it dynamically connected}
part of the pore space in porous media \cite{banavar87,kostek92}. 
Following Ref.~\cite{kostek92}, we write $\Lambda$
in terms of the permeability and tortuosity of the system,
\begin{equation}
\label{lambda}
k = \frac{\Lambda^2}{12~T^2}
\end{equation}
Note that defining a characteristic
length, as the ratio between pore volume $V$ and surface area $S$
would yield $V/S = H/T \neq \Lambda$, which does not 
depend on the effects exerted on the fluid flow by 
the complex  geometry of the fracture.

Now, assuming that we have a winding channel of length $L_e$,
effective aperture for convective transport $\Lambda$ 
and actual aperture $ A = H / T$ (given
by volume conservation), we may apply the same reasoning as in Section
\ref{taylor2D} to get the dispersion coefficient.
After replacing $H$ by the width of the effective channel $H/T$,
$U$ by the mean velocity in the convective part of the channel
$U H / \Lambda$, and taking into account the tortuosity,
we obtain
\begin{equation}
\label{taylor_lambda}
D_{\parallel} = \frac{D_m}{T^2} 
\left( 1 + \frac{Pe^2}{210} \left(\frac{H}{\Lambda T}\right)^2 \right)
\end{equation}

Thus, to see how well the $\Lambda$ parameter can be used to estimate
the dispersion, we will compare the values obtained both
from dispersion measurements (Eq.~(\ref{taylor_lambda})) 
and from the permeability (Eq.~(\ref{lambda})).

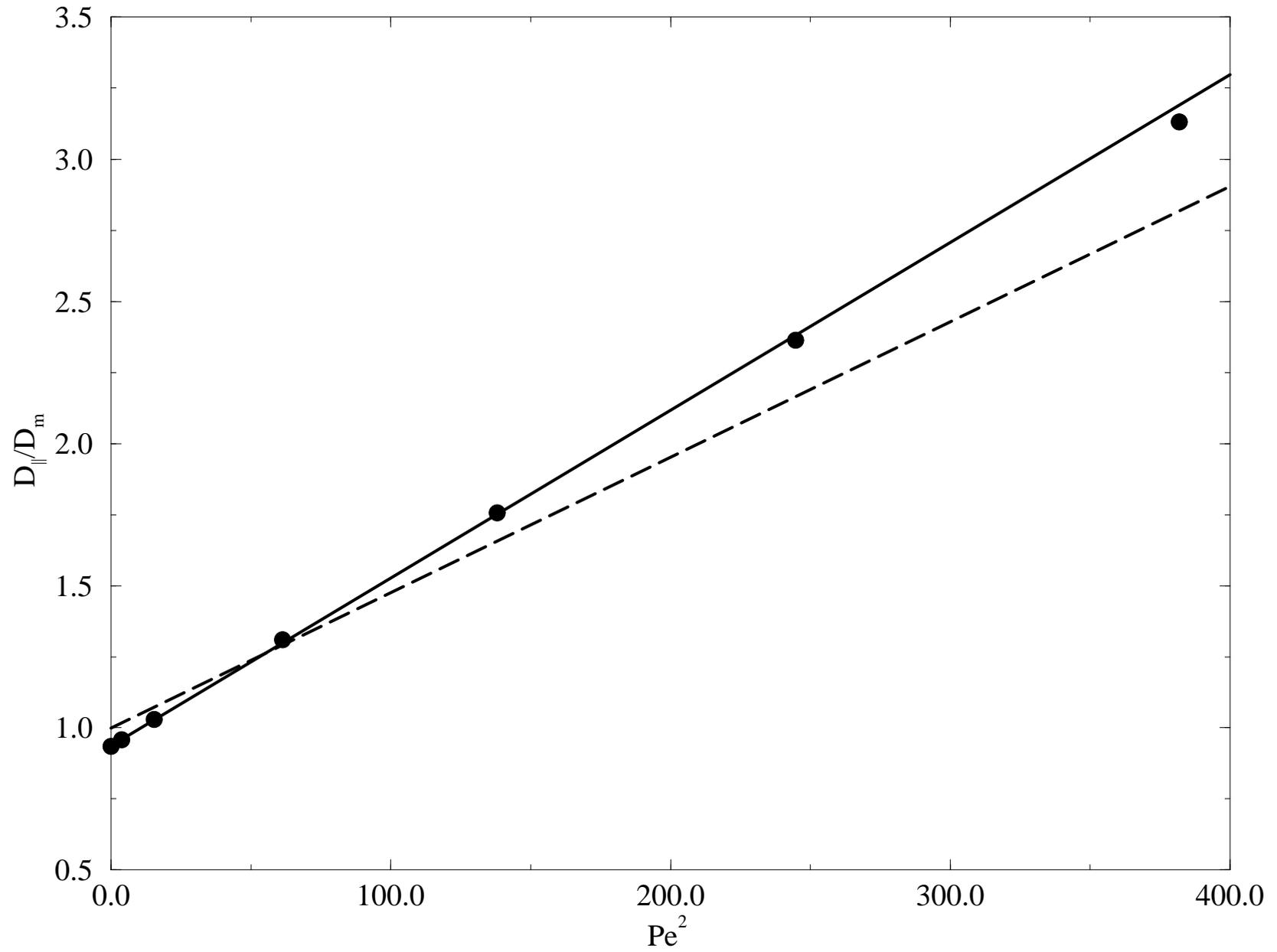
\begin{figure}
\caption[T]{Dispersion coefficient as a function of the 
P\'eclet number squared. Solid line corresponds to the best fit using 
Eq.~(\ref{taylor_lambda}), with $D_m$ and $\Lambda$ as 
adjustable parameters. Dashed line corresponds to Taylor dispersion in 
a straight channel with equal aperture $H=16$.}
\label{no_shift}
\end{figure}

In Figure \ref{no_shift} we show the numerical results obtained for the
dispersion coefficient when varying the injection rate.
The linear behavior shows that, as expected,
Taylor dispersion is governing tracer spreading. 
The fracture gap is $H=16$, the length of the system is $L=512$, and
the relaxation parameter used is $\lambda_D=1.9$. 
The molecular diffusivity and the lambda parameters obtained from the
the best fit are: $D_m = (8.86 \pm 0.01) \times 10^{-3}$ 
which differs by only $1\%$ from the theoretical value; and
$\Lambda= 13.4\pm0.1$. On the other hand, from the flow rate computed
in numerical simulations, and estimating the permeability 
by Eq.~(\ref{lambda}), we get $\Lambda=15.3\pm0.2$. 

We now turn to shifted surfaces. In the presence of a small 
lateral shift $d$ between complimentary surfaces, 
most of the previous discussion remains valid.
The local aperture now varies with position and
$H$ is the average aperture of the channel.
The difference in height between 
surfaces at any point $x$ along the channel is of order 
$d^\zeta$, while the vertical excursion of the fracture
between points separated a distance $\Delta x$ is $(\Delta x)^\zeta$.
Therefore, at large length scales, $\Delta x \gg d$, 
the fracture may be considered as a winding channel
of length $L_e$ and effective aperture $\Lambda$ and
where the ratio between the length of the channel $L_e$ 
and the system length $L$ is the tortuosity factor.

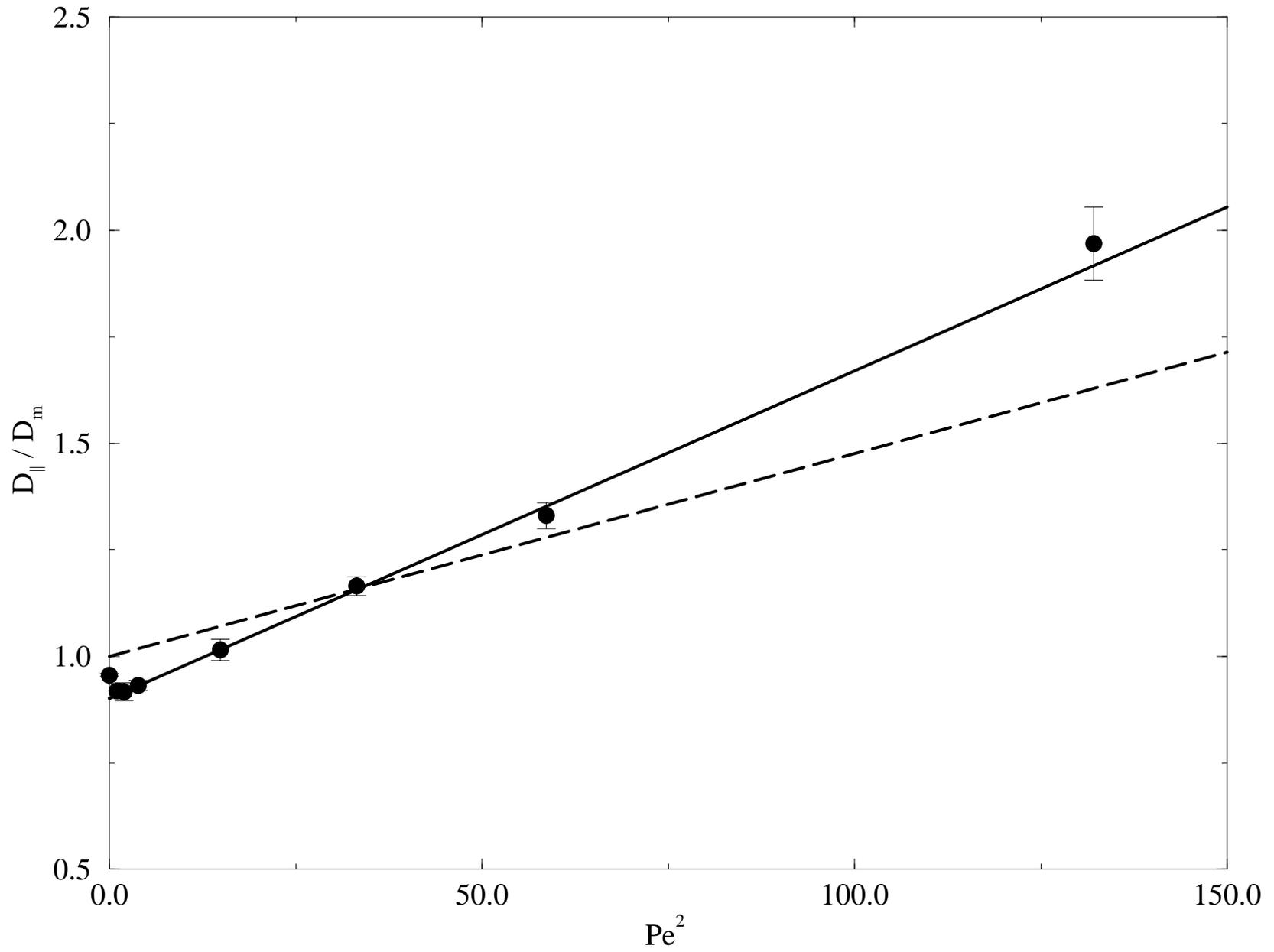
\begin{figure}
\caption[D]{Dispersion coefficient as a function of the 
P\'eclet number squared. Solid line corresponds to the best fit using 
Eq.~(\ref{taylor_lambda}), with $D_m$ and $\Lambda$ as 
adjustable parameters. Dashed line corresponds to Taylor dispersion in 
a straight channel with equal aperture $H=16$.}
\label{shift}
\end{figure}

In Figure \ref{shift} we show the dispersion coefficient, 
as a function of the P\'eclet number. 
The two fracture surfaces are vertically displaced by $H=16$ and
laterally shifted by $d=8$. As expected, 
we observed a linear dependence, in agreement with 
Taylor-like spreading. From the best fit of the numerical results
we obtain $D_m = (8.5 \pm 0.1) \times 10^{-3}$ which differs
less that $3\%$ from the theoretical one ($\lambda_D=1.9$);
and $\Lambda= 14.6\pm0.2$. Whereas, by means of flow rate data
computed in numerical simulations and Eq. (\ref{lambda})
for the permeability, we obtain $\Lambda= 11.7\pm0.5$.

Even though  estimated values of $\Lambda$ 
are in fairly good agreement (within a $20\%$ discrepancy),
it is also clear that the $\Lambda$-parameter 
fails to completely predict 
the enhancement of dispersion due to the complex 
geometry of the fractures.
Nevertheless, we believe that the presence of low-velocity zones,
is the only possible feature that
accounts for the enhancement in 
the spreading of tracer particles.

Let us note that the uncertainty in the computed 
dispersion coefficient is considerably larger than in the case 
where there is no lateral shift. 
It has been shown, in three-dimensional fractures
and under lubrication approximation for the velocity field,
that a lateral shift yields geometric dispersion
for tracer particles advected along the flow \cite{roux98}.
This effect is due to different mean velocity along 
different streamlines. In our case, the two-dimensional 
nature of the fractures
prevents the presence of geometric dispersion, given that 
the height-averaged velocity is constant throughout
the system. However, an analogous effect appears
when averaging over different fractures,
giving rise to a large dispersion in the computed mean velocity 
and dispersion coefficient. In fact, from the previous 
discussion an uncertainty proportional to the
mean velocity is expected when averaging the
dispersion coefficient (corresponding to a geometric 
dispersion term in fully three-dimensional fractures).

In order to validate the dispersion results
obtained by means of the LB method, 
we shall now compare them with results computed
via a Monte-Carlo (MC) approach to dispersion process.
In the MC method, one follows the
displacement of a large number of particles, or random walkers,  
moving in a two-dimensional fracture. The motion of each particle 
is a combination of the effects of molecular diffusion 
and convection (we assume, as in the LB simulations
that tracer particles moves independently of each other). 
In time $\Delta t$, a particle is displaced according to
\begin{equation}
\Delta {\bf x} = {\bf u}({\bf x}) \Delta t + {\bf \hat n}
(4 D_m \Delta t)^{1/2}
\label{randomstep}
\end{equation}
where ${\bf u}({\bf x})$ is the velocity field obtained by the
LB method, ${\bf \hat n}$ is a unit vector with
random orientation, and the amplitude of the random steps
has been chosen so that the variance of any coordinate is
$2 D_m \Delta t$\cite{kurowski94,koplik94,zhang97}.
Boundary condition at solid surfaces are implemented as in 
\cite{kurowski94}, where those random steps that would take the 
particle outside the channel are suppressed. 
The sequence of steps is repeated while recording the 
distance $\Delta x$ from the initial position 
of the particle. The process is repeated
for a large number of particles and average values are computed.

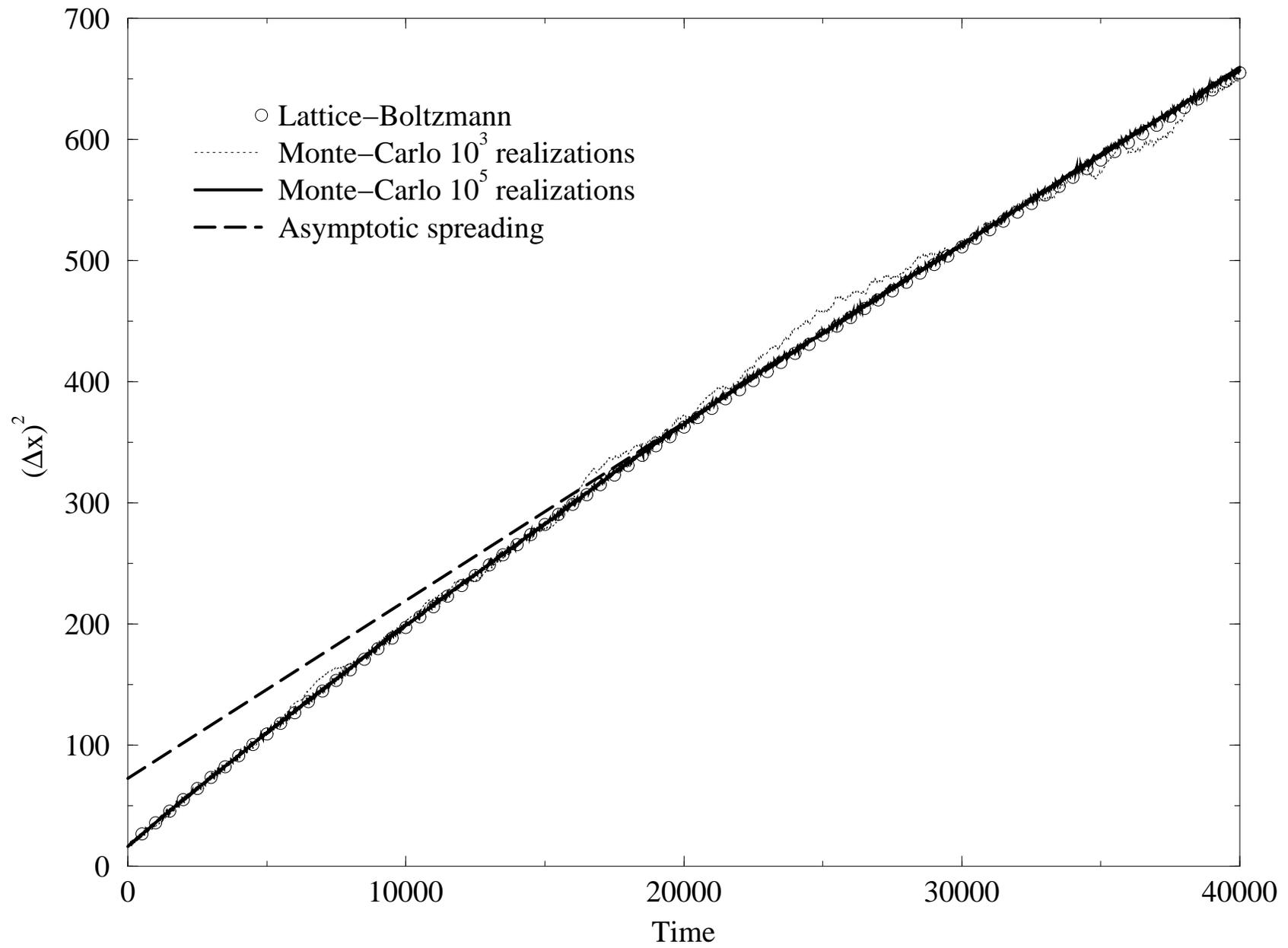
\begin{figure}
\caption[montecarlo]{Mean square displacement of tracer particles as a function of time, 
in a single fracture. (Mean aperture $H=16$; Shift between surfaces $d=8$; 
Length $L=512$. Mean velocity $U\approx10^{-3}$. $D_m=0.008771$.).
Solid line and open circles correspond to numerical 
MC and LB simulations respectively. 
Dashed line is the best linear fit for large times, i.e. the
asymptotic linear spreading of the tracer ($D=7.4 \times 10^{-3}$).}
\label{montecarlo}
\end{figure}

We found an excellent agreement between the two methods.
In Fig. \ref{montecarlo} we compare the mean square displacement 
as a function of time obtained using LB and MC
methods. The agreement is evident, and it can be seen
that after releasing $10^5$ particles in the MC simulations, 
the noise is negligible.
Both simulations corresponds to a fracture of mean
aperture $H=16$, length $L=512$, and a lateral shift 
between surfaces $d=8$. It is also interesting to note,
in Fig. \ref{montecarlo}, a change in the slope at time 
$t \sim 20000$, which approximately corresponds
to the characteristic time
for transverse diffusion across the aperture 
$\tau_D \approx H^2/2D_m \sim 15000$. Therefore, 
this marked change in slope is showing the transition
towards Taylor-like spreading at times larger than $\tau_D$.


\subsection{Dispersion at small P\'eclet numbers}
\label{smallPe}

In a straight channel, the presence of convection increases the 
dispersion of tracer particles. The same effect is usually 
found in porous media, including fracture systems. 
However, in the case where surfaces are laterally shifted, 
and at very small P\'eclet number, we observed the opposite behavior.
That is, the presence of convective transport inside the fractures
reduces the dispersion of tracer particles. This effect can be
observed in Fig.~\ref{shift}, at very small Pe  
the dispersion coefficient grows as the flow rate decreases.
In order to validate this observation we 
simulate the dispersion process with two other methods. 
The first one is the MC method presented before. 
The second method is a variation of the MC where, 
instead of using the velocity field computed by means 
of LB simulations, we use lubrication approximation 
for velocities. In the lubrication approximation the
velocity $u(x,z)$ is given by,
\begin{equation}
\label{lubrication}
u(x,z) = \frac{6 Q}{H^3(x)}~z~(H(x)-z)
\end{equation}
where $H(x)$ is the local aperture of the fracture.
 
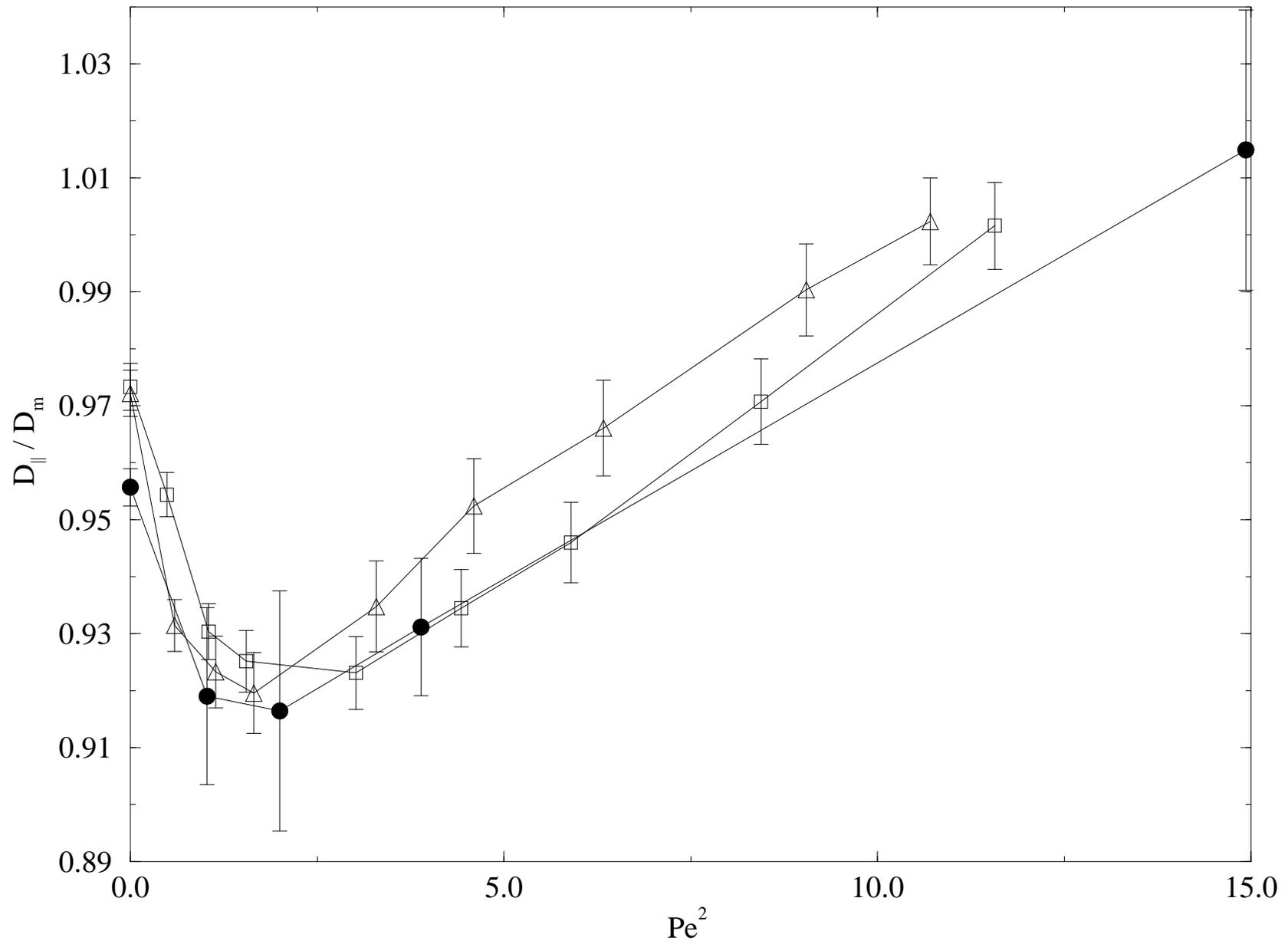
\begin{figure}
\caption[L]{Dispersion coefficient at low P\'eclet numbers. 
Different results correspond to different simulation methods.
Solid circles correspond to LB method.
Squares correspond to MC simulations where the velocity
was computed using LB. Triangles correspond to MC
simulations, using the velocity field given by the
lubrication approximation.}
\label{low}
\end{figure}

In Figure \ref{low} we show the results obtained using the three different 
methods in a range of Pe from $0$ to $4$. The agreement between
methods is excellent and all show an initial 
decrease in dispersion due to the presence of weak convective transport. 
We believe that, at very low velocities, the main effect of convective 
transport is to carry out tracer from {\it low-velocity}
zones, and thus, reducing dispersion. This explanation 
is supported by the fact that lubrication
approximation yields very similar results. 
(Note that this approximation assumes local Poiseuille 
profile without accounting for possible stagnant zones). 
Unfortunately, the small P\'eclet numbers at which this surprising 
effect arises, makes it very difficult to be
experimentally measured. 


\section{Conclusions}
\label{conclusions}

In first place, we have presented a new set of boundary conditions
to describe diffusive transport within the lattice-Boltzmann
method. We tested this new boundary rule in two situations, 
simple diffusion and Taylor dispersion in two-dimensional
straight channels. We showed that the proposed bounce-forward rule
improves the accuracy of the method and does not possess  the 
undesirable effects of the bounce-back rule, 
i.e., dependence of the diffusion coefficient on the aspect ratio
of a straight channel and transient concentration gradients
near solid walls. Even though the accuracy might be recovered
in the BB case using small diffusivities, we showed that this option
leads the simulations towards the numerical 
instability border. 

We then turn to study diffusive transport in two-dimensional
self-affine fractures. 
First, we showed that the slow-down in diffusive transport
can be accounted for by the purely-geometric tortuosity factor.
Our numerical simulations have verified all the implications of 
this result, that is, linear spreading on time, Gaussian 
distribution of tracer and independence of geometrical effects on
the actual value of the diffusivity. Secondly, 
using analytic arguments in the limit of small aperture 
fluctuations, we have obtained an expression for
the tortuosity in terms of the fracture gap and the Hurst
exponent characterizing the fracture surface.  
Numerical simulations verify the validity of the 
theoretical approach even when the discrete 
nature of the surface seems to affect the asymptotic 
scaling law. 

Finally we studied tracer dispersion in fractures in the 
presence or not of a lateral shift between 
complementary surfaces. In both cases we showed that 
tracer spreading can be described as analogous to Taylor-like 
dispersion in a straight channel.
In the case without lateral shift we showed that,
the enhancement in dispersion can be mostly 
understood assuming an effectively
reduced aperture for fluid transport, due to
the rugosity of the surface. The $\Lambda$-parameter
measuring this effective aperture, and computed
from permeability measurements, was shown to be
similar to that estimated from dispersion measurements. 
We also presented a novel result, showing a decrease in
tracer dispersion when convective transport is set in
the fracture. This last observation was obtained
in the frame of three different numerical approaches.

\section*{Acknowledgments}
We thank J. P. Hulin, F. Plourabou\'e for many fruitful discussions and
M. Tanksley, N. Rakotomalala and D. H. Rothman for useful discussions 
on the lattice-Boltzmann method. G. D. 
thanks M. Tirumkudulu and I. Baryshev for their helpful comments.
G. D. was partially supported by CONICET Argentina and The
University of Buenos Aires. This research was supported by 
the Geosciences Research Program, Office of
Basic Energy Sciences, U.S. Department of Energy, and computational facilities
were provided by the National Energy Resources Scientific
Computer Center.

\newpage

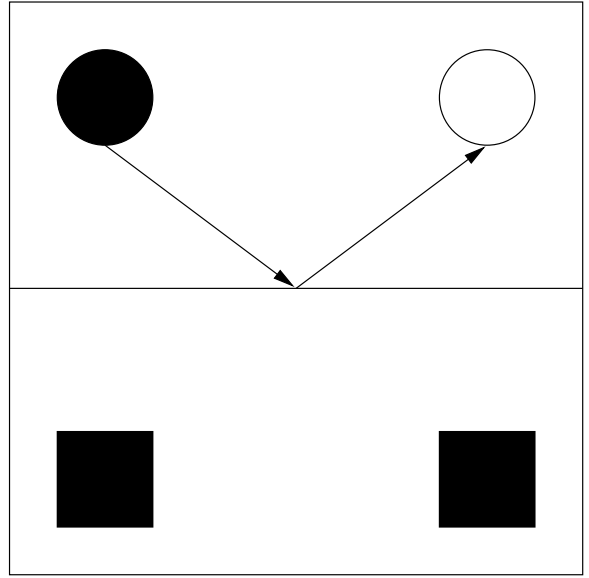
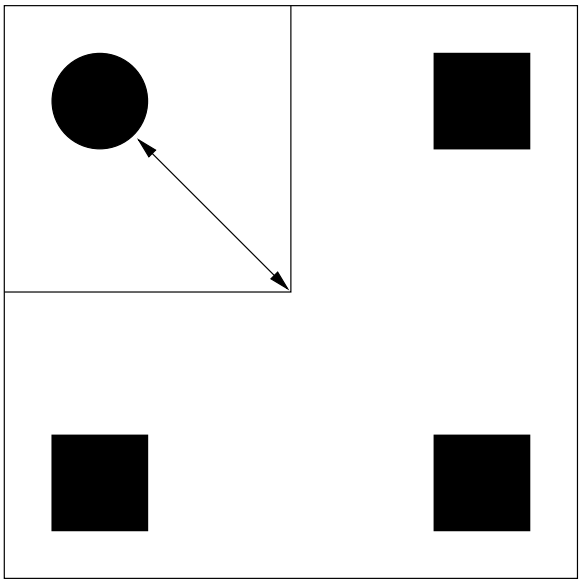
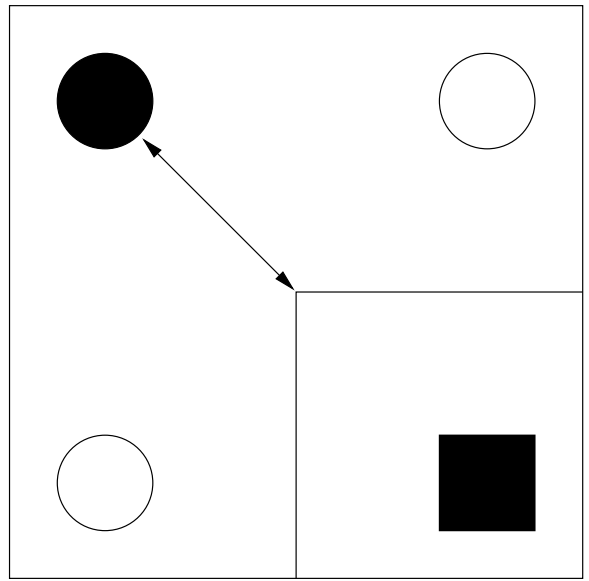
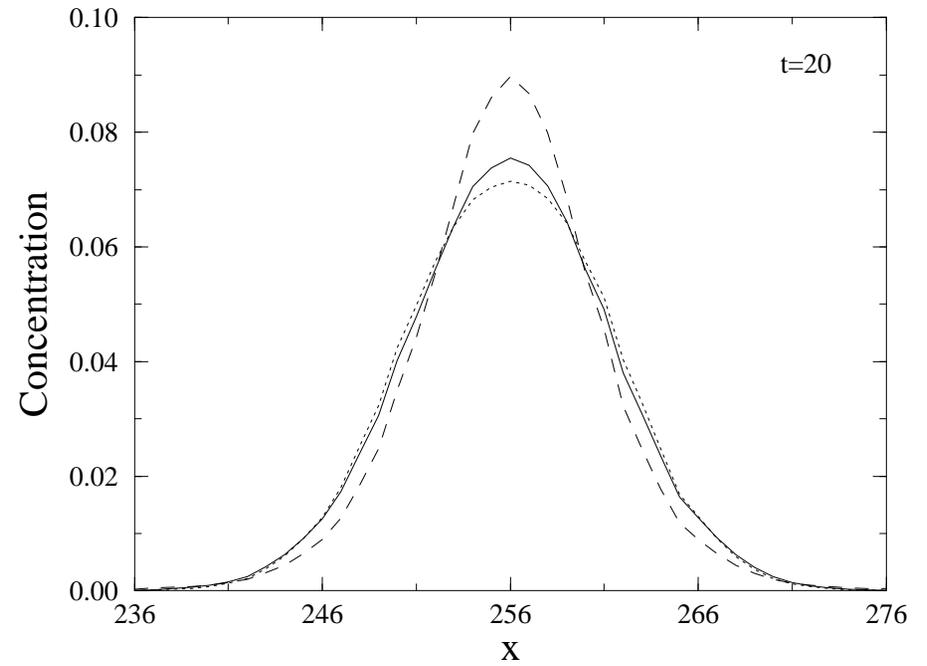
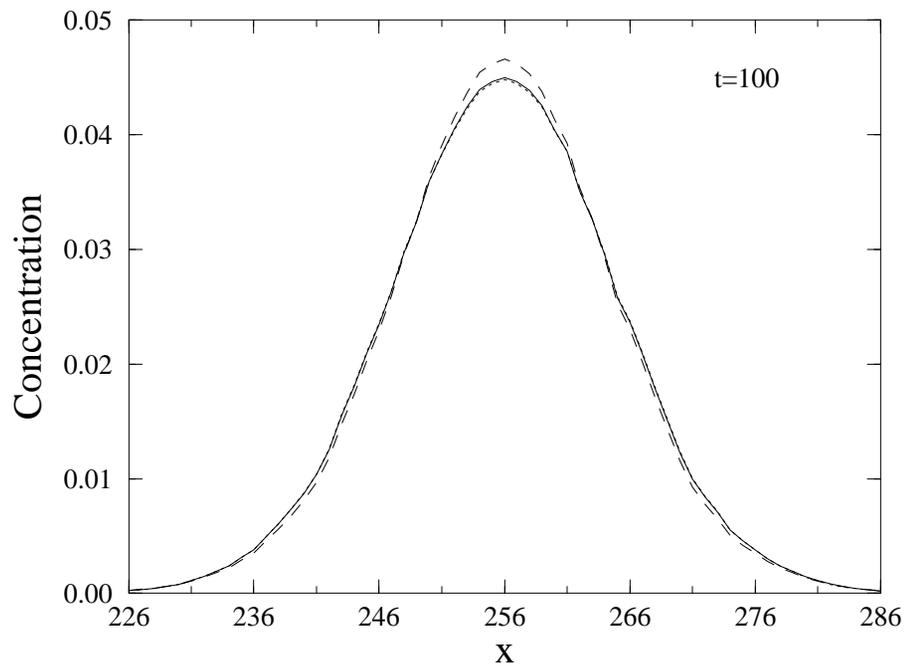
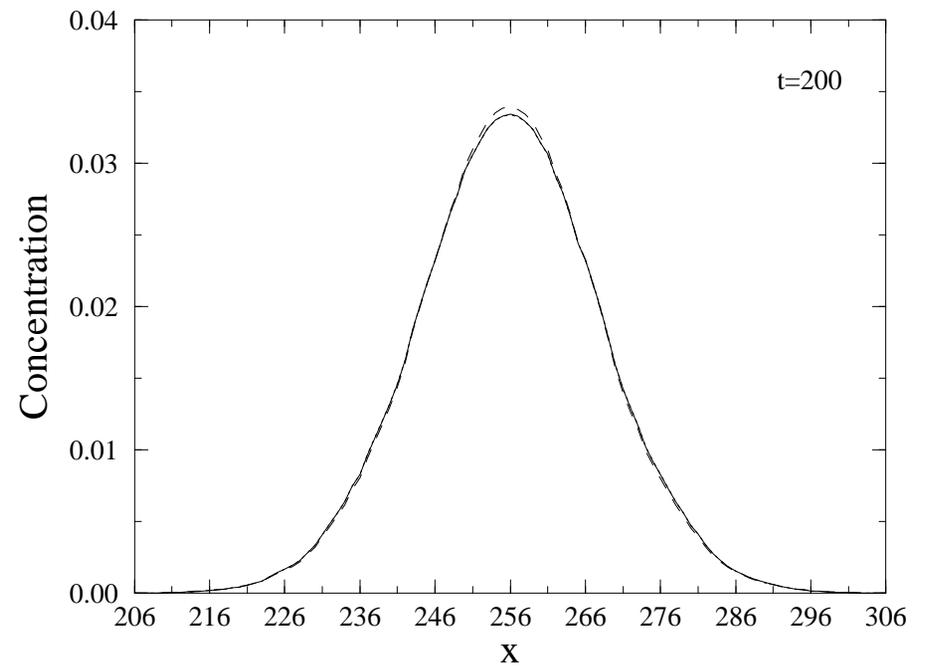
\end{document}